\documentclass[onecollarge]{svjour2}       
\smartqed  
\usepackage{graphicx}
\usepackage{hyperref}

\usepackage{subfigure}

\usepackage{amsmath}

\journalname{Journal of Statistical Physics}

\begin{document}
\title{Optimal segmentation of directed graph and the minimum number of feedback arcs}

\titlerunning{Optimal segmentation and minimum feedback arc set}

\author{Yi-Zhi Xu$^{1,2}$ \and Hai-Jun Zhou$^{1,2}$}

\institute{
  $^1$Key Laboratory of Theoretical Physics, Institute of Theoretical Physics,
  Chinese Academy of Sciences, 
  Zhong-Guan-Cun East Road 55, Beijing 100190, China 
  (\email{zhouhj@itp.ac.cn}) \\
  $^2$School of Physical Sciences, University of Chinese Academy of Sciences,
  Beijing 100049, China
}

\date{16 May, 2017}

\maketitle

\begin{abstract}
  The minimum feedback arc set problem asks to delete a minimum number of arcs (directed edges) from a digraph (directed graph) to make it free of any directed cycles. In this work we approach this fundamental cycle-constrained optimization problem by considering a generalized task of dividing the digraph into $D$ layers of equal size. We solve the $D$-segmentation problem by  the replica-symmetric mean field theory and belief-propagation heuristic algorithms. The minimum feedback arc density of a given random digraph ensemble is then obtained by extrapolating the theoretical results to the limit of large $D$. A divide-and-conquer algorithm (nested-BPR) is devised to solve the minimum feedback arc set problem with very good performance and high efficiency. \\
\\

\end{abstract}

\maketitle

\section{Introduction}

\begin{figure}
  \begin{center}
    \includegraphics[width=0.5\linewidth]{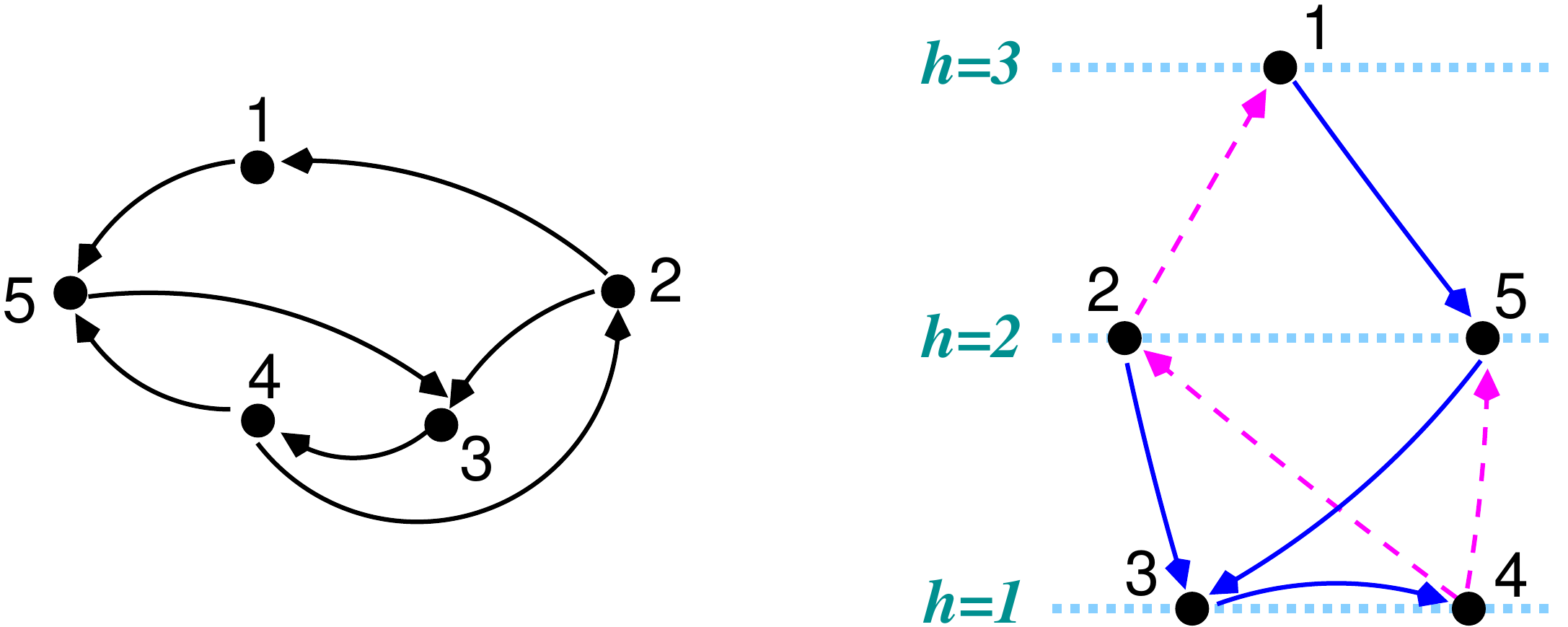}
  \end{center}
  \caption{
    \label{fig:Dsegment}
    An illustration of the digraph $D$-segmentation problem. (left) A digraph example $G$ containing $N=5$ vertices and $M=7$ arcs. (right) This digraph is partitioned into $D=3$ layers with three feedback arcs (the dashed arrows), and the states of the vertices are $h_3=h_4=1$, $h_2=h_5=2$, and $h_1=3$.
  }
\end{figure}

Directed graphs (digraphs) are structural descriptions for various many-body systems whose constituent elements are related by directional interactions. The neural network of the brain and the gene regulation network of the cell are prominent examples of digraphs. The edges of a digraph have directions, and these directed edges are referred to as arcs to distinguish them from the edges of undirected graphs. An arc $(i, j)$ connects two vertices and it points from vertex $i$ to vertex $j$. Usually there are many arcs in a digraph. In this work we shall be interested in sparse digraphs whose total number $M$ of arcs is proportional to the total number $N$ of vertices. The arc density $\alpha$ of a sparse digraph, with $\alpha \equiv \frac{M}{N}$, is then bounded by a constant value even for $N \rightarrow \infty$. A vertex in a sparse digraph has on average $\alpha$ in-coming arcs which point to it and $\alpha$ out-going arcs which point from it. When the arc density $\alpha$ is greater than unity, many directed cycles are expected to form in the digraph. A directed cycle is a closed path of arcs. For example, in Fig.~\ref{fig:Dsegment} the directed path formed by arcs $(1, 5)$, $(5, 3)$, $(3, 4)$, $(4, 2)$, and $(2, 1)$ is a directed cycle of length five connecting all the vertices of the digraph.

An issue of fundamental importance in digraph research is the minimum feedback arc set (FAS) problem, which is also encountered in many practical applications, such as inferring the hierarchical structure of the digraph \cite{Lan-Mezic-2011,Gupte-etal-2011,Xu-Lan-2015,Zhao-Zhou-2016b} and detecting the dominant direction of information flow \cite{Ispolatove-Maslov-2008,DominguezGarcia-Pgolotti-Munoz-2014}. The minimum FAS problem aims at breaking all the directed cycles of an input digraph by deleting as few arcs as possible. The deleted arcs are classified as feedback arcs and they form a minimum FAS, while all the remaining arcs are classified as feedforward arcs. The subgraph formed by all the vertices and all the feedforward arcs is free of any directed cycles, it is a directed acyclic graph admitting the maximum number of arcs. It should be emphasized that a FAS is a collective property of the digraph, and a feedback arc must be understood as a member of such a set.

The minimum FAS problem belongs to the class of non-deterministic polynomial hard (NP-hard) combinatorial optimization problems \cite{Karp-1972}. It is very likely that this problem can not be exactly solved by any polynomial-time algorithm. Various heuristic procedures have been investigated to solve it approximately but efficiently \cite{Lan-Mezic-2011,Gupte-etal-2011,Xu-Lan-2015,Ispolatove-Maslov-2008}. In a recent work \cite{Zhao-Zhou-2016b}, Zhao and one of the present authors proposed a spin glass model for this cycle-constrained optimization problem and derived a mean field theory for estimating the  minimum number of feedback arcs and for constructing near-minimum feedback arc sets. The physics-inspired message-passing algorithm performs slightly worse in comparison with a simulated annealing (SA) algorithm adapted from \cite{Galinier-Lemamou-Bouzidi-2013}.

In the present paper we continue to study the minimum FAS problem as a statistical physical system, but we look at it from a different angle. We introduce the optimal segmentation problem, which asks to evenly distribute the $N$ vertices of a digraph to $D$ layers under the constraint that the total number of arcs pointing from lower layers to higher layers should be minimized. The minimum FAS problem is a limiting case of this more general $D$-segmentation  problem. This new perspective enables us to improve the computation on the minimum size of feedback arc sets for different ensembles of random digraphs, and it also brings new algorithmic insights on more efficient ways of tackling the minimum FAS problem. We are indeed satisfied to find that the nested-BPR algorithm inspired by this new theoretical approach beats SA both in performance and in speed. 

The optimal $D$-segmentation problem can be regarded as a natural extension of the graph partitioning problem, which is one of the first combinatorial optimization problems studied in the statistical physics community \cite{Fu-Anderson-1986}. It is an interesting graph optimization task on its own sake. The mean field theory of this paper can be used to estimate the minimum number of arcs pointing from lower layers to higher layers, and the associated message-passing algorithms achieve nearly-optimal solutions for single digraph instances. We expect our methods to be useful in studying the structural and dynamical properties of various digraphs.

The next section defines the theoretical model of digraph $D$-segmentation and establishes the important link with the minimum FAS problem. Section~\ref{sec:RS} contains the replica-symmetric mean field theory, and Sec.~\ref{sec:result} reports some theoretical results obtained on three types of random digraphs. The algorithmic applications of the mean field theory are then discussed in Sec.~\ref{sec:algorithm}. We conclude this work in Sec.~\ref{sec:conclusion} and make some further discussions.

\section{Model}
\label{sec:model}

Given a digraph $G$ formed by $N$ vertices and $M$ arcs, a $D$-segmentation of $G$ is simply a partitioning of the $N$ vertices into $D$ layers such that the number of vertices in each layer is the same (in the case of $N\, {\rm mod}\, D =0$) or differs by at most one (in the case of $N\, {\rm mod}\, D \neq 0$), see Fig.~\ref{fig:Dsegment}. We can assign an integer height $1 \leq h \leq D$ to each layer to distinguish the $D$ different layers, then each vertex $i$ of the digraph has a state $h_i \in \{1, 2, \ldots, N\}$. A configuration of the digraph is then $\underline{h} \equiv (h_1, h_2, \ldots, h_N)$. Let us decompose the vertex number as $N= D n + r$ with $n$ being the largest integer satisfying $D n \leq N$ and $r \geq 0$ being the remainder. For $\underline{h}$ to be a valid $D$-segmentation configuration, it needs to satisfy the following uniformity condition:
\begin{equation}
  \label{eq:uniformity}
  \sum\limits_{i=1}^{N} \delta_{h_i}^{h} = \left\{ \begin{array}{ll}
    n & \quad \quad \quad (r < h \leq D )\; , \\
    & \\
    n+1  & \quad \quad \quad (1\leq h \leq r) \; ,
  \end{array}
  \right.
\end{equation}
where $\delta_{h^\prime}^{h}$ is the Kronecker symbol such that $\delta_{h^\prime}^{h}=1$ if $h=h^\prime$ and $\delta_{h^\prime}^{h}=0$ if $h\neq h^\prime$. If $N$ is divisible by $D$ (so $r=0$), the condition (\ref{eq:uniformity}) means that every layer contains the same number $n$ of vertices; if $r>0$, Eq.~(\ref{eq:uniformity}) means that each of the lowest $r$ layers contains one more vertex than each of the remaining $(D-r)$ layers does.

With respect to a valid $D$-segmentation configuration $\underline{h}$, an arc $(i, j)$ will be considered as a feedback arc if and only if $h_i < h_j$, namely the arc points from a vertex in a lower layer to another vertex in the higher layer. The total arc energy of a valid configuration $\underline{h}$ is defined as the total number of feedback arcs in this configuration. We should emphasize that, different from the model introduced in \cite{Zhao-Zhou-2016b}, a horizontal arc between two vertices of the same layer costs no energy in the $D$-segmentation problem. An optimal solution for the $D$-segmentation problem is a valid configuration $\underline{h}$ whose arc energy achieves the global minimum value. Let us denote this minimum arc energy as $R_0(D)$ and define the minimum feedback arc density as $\rho_0(D) \equiv \frac{R_0(D)}{M}$.

The digraph $D$-segmentation problem is closely related to the partitioning problem of an undirected graph \cite{Fu-Anderson-1986}, whose objective is to split a graph into two disconnected parts of comparable sizes by cutting the minimum number of edges. This later problem has been extensively investigated by the replica and the cavity method of statistical physics (see, e.g., \cite{Mezard-Parisi-1987,Sherrington-Wong-1987,Lai-Goldschmidt-1987,Sulc-Zdeborova-2010,Kawamoto-Kabashima-2015}). A major new feature of the digraph case is that not all the arcs between two layers need to be deleted but only those upward arcs from the lower layer to the higher layer. When $D>2$, the state space of the $D$-segmentation problem is also much larger as each vertex can choose among $D$ different states.

\begin{figure}
  \begin{center}
    \includegraphics[width=0.5\linewidth]{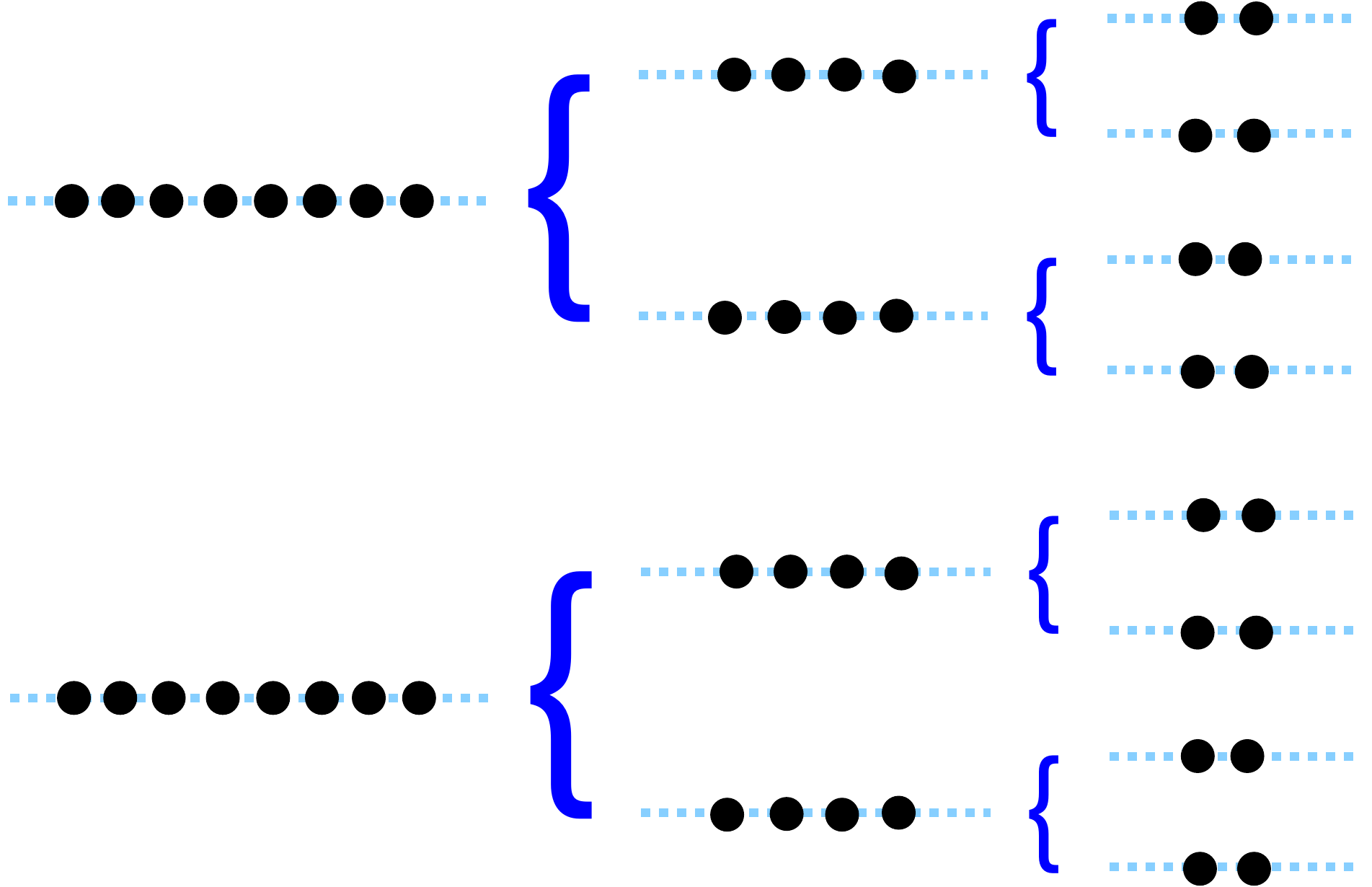}
  \end{center}
  \caption{
    \label{fig:multiple}
    An illustration of a series of $D$-segmentation problems with $D=2^m$ (here only $m=1, 2, 3$ are shown).
  }
\end{figure}

At the thermodynamic limit $N\rightarrow \infty$, the minimum feedback arc density $\rho_0(D)$ of a digraph $G$ has the following nice monotone property, namely
\begin{equation}
  \label{eq:rho0}
  \rho_0(D) \leq \rho_0(2 D) 
\end{equation}
for any $D \geq 2$. This property is a direct consequence of the many-to-one mapping from solutions of the $2 D$-segmentation problem to those of the $D$-segmentation problem (Fig.~\ref{fig:multiple}). For each solution $\underline{h}$ of the $2 D$-segmentation problem, an solution $\underline{h}^\prime$ of the $D$-segmentation problem can be obtained simply by merging the vertices of two consecutive layers of $\underline{h}$ into a single layer of $\underline{h}^\prime$. The number of feedback arcs decreases during this merging process because the arcs within each merged layer can not be feedback arcs. Therefore the minimum arc energy $R_0(D)$ of the $D$-segmentation problem must not exceed the minimum arc energy $R_0(2 D)$ of the $2 D$-segmentation problem, and then the inequality (\ref{eq:rho0}) follows.

For a very large digraph $G$ with $N\rightarrow \infty$ and constant arc density $\alpha$, because the minimum feedback arc densities
$$
\rho_0(2),\ \rho_0(4),\ \ldots,\ \rho_0(2^m),\ \ldots
$$
form a monotone sequence and $\rho_0(D)$ is upper-bounded by $\rho_0(D) < 1$, a well-defined limiting value $\rho_\infty$ must be reached by $\rho_0(D)$ at $D\rightarrow \infty$. This limiting value $\rho_\infty$ must be identical to the fraction of arcs (with respect to the total arc number $M$) in a minimum FAS of the digraph $G$, because  at $D\rightarrow \infty$ (or  $D \sim N$) each minimum $D$-segmentation solution must correspond to a minimum FAS of $G$.

In the present paper we will exploit this very significant relationship to determine the arc density ($\rho_\infty$) of the minimum FAS problem and to solve the minimum FAS problem efficiently for single digraph instances.
 
\section{Replica-symmetric mean field theory}
\label{sec:RS}

The global constraints (\ref{eq:uniformity}) require to count exactly the number of vertices in each integer layer $h\in [1, D]$. In the mean field theory these constraints are imposed in the average sense by introducing on each layer $h$ an occupation cost $C(h)$. The total energy $E(\underline{h})$ for a microscopic configuration $\underline{h}=(h_1, h_2, \ldots, h_N)$ of the digraph $G$ is then defined as
\begin{equation}
  E(\underline{h}) = \sum\limits_{i=1}^{N} C(h_i) + 
  \sum\limits_{(j, k) \in G} E_{j k}(h_j, h_k) \; .
  \label{eq:EofModel}
\end{equation}
In the above expression the energy of an arc $(j, k)$ is $E_{j k}(h_j, h_k) = 0$ if $h_j \geq h_k$ and $E_{j k}(h_j, h_k)=1$ if $h_j < h_k$. The partition function $Z$ corresponding to Eq.~(\ref{eq:EofModel}) is
\begin{equation}
  Z \equiv \sum\limits_{\underline{h}} e^{-\beta E(\underline{h})}
  = \sum\limits_{\underline{h}}
  \prod\limits_{i=1}^{N} e^{- \beta C(h_i)} \prod\limits_{(j, k)\in G}
  \psi_{j k}(h_j, h_k) \; ,
  \label{eq:Zexp}
\end{equation}
where $\beta$ is the inverse temperature. The Boltzmann weight of an arc $(j, k)$ is $\psi_{j k}(h_j, h_k) = 1$ for $h_j \geq h_k$ and $\psi_{j k}(h_j, h_k) = e^{-\beta}$ for $h_j < h_k$.

Notice that the summation in Eq.~(\ref{eq:Zexp}) is over all the $D^N$ microscopic configurations. To ensure that the partition function is contributed predominantly by the valid occupation configurations, the occupation cost function $C(h)$ has to be carefully adjusted \cite{Sulc-Zdeborova-2010}. Let us denote by $q_i^{h}$ the marginal probability that vertex $i$ is staying at layer $h$. The averaged total number of vertices at layer $h$ is then $\sum_{i=1}^{N} q_i^h$. Equation (\ref{eq:uniformity}) then leads to the constraint of
\begin{equation}
  \label{eq:hcondition}
  \frac{1}{N} \sum\limits_{i=1}^{N} q_i^{h} = \frac{1}{D}
  \; ,  \quad \quad \quad \forall\  h\in\{1, 2, \ldots, D\} \; .
\end{equation}
This equation means that the fraction of vertices at each layer $h$ should be equal to $\frac{1}{D}$. The cost function $C(h)$ can be uniquely determined by Eq.~(\ref{eq:hcondition}) up to an unimportant constant term.

We now develop the replica-symmetric (RS) mean field theory for the digraph $D$-segmentation problem based on the Bethe-Peierls approximation \cite{Mezard-Parisi-2001,Yedidia-Freeman-Weiss-2001,Yedidia-Freeman-Weiss-2005,Mezard-Montanari-2009}.  For readers with no background knowledge on statistical mechanics, let us point out that the  RS mean field theory for such a $D$-state graphical system can also be derived through the mathematical trick of graph loop expansion \cite{Xiao-Zhou-2011,Zhou-Wang-2012,Mori-2015}.

\subsection{Belief propagation equation}

Let us denote by $p(j)$ the set of parent vertices of vertex $j$, namely $p(j) \equiv \{i : \, (i, j)\in G\}$; and similarly by $c(j) \equiv \{k  : \, (j, k)\in G\}$ the set of child vertices of vertex $j$. Let us assume that the states of all the vertices in the set $p(j)$ and $c(j)$ are mutually independent of each other in the absence of vertex $j$. Under this assumption of conditional independence, then in the absence of vertex $j$, the joint state distribution of all the nearest neighboring vertices of $j$ has the following factorized form 
$$
\prod\limits_{i\in p(j)} q_{i\rightarrow j}^{h_i}
\prod\limits_{k\in c(j)} q_{k\rightarrow j}^{h_k}
\; . 
$$
The quantities $q_{i\rightarrow j}^{h_i}$ and $q_{k\rightarrow j}^{h_k}$ in the above expression are, respectively, the marginal state distribution of vertex $i\in p(j)$ and vertex $k\in c(j)$ in the absence of vertex $j$. They are referred to as cavity probability distributions in the literature. Because vertex $j$ interacts with all its nearest neighboring vertices, the marginal probability distribution $q_{j}^{h_j}$ is then expressed as
\begin{equation}
  \label{eq:qimarginal}
  q_j^{h_j} =   \frac{1}{z_j} e^{-\beta C(h_j)} w_j(h_j) \; ,
\end{equation}
where the statistical weight $w_j(h_j)$ is expressed as
\begin{equation}
  w_j(h_j)  \equiv  \prod\limits_{i\in p(j)}
  \Bigl[e^{- \beta} + (1- e^{-\beta})
    \sum\limits_{h_i = h_j}^{D} q_{i\rightarrow j}^{h_i} \Bigr]
  \prod\limits_{k \in c(j)} \Bigl[e^{-\beta} + (1-e^{-\beta}) 
    \sum\limits_{h_{k} = 1}^{h_j} q_{k \rightarrow j}^{h_{k}}
    \Bigr] \; ,
  \label{eq:wj}
\end{equation}
and $z_j$ is the probability normalization constant determined by
$z_j = \sum_{h=1}^{D} e^{-\beta C(h)} w_j(h)$.

By applying the same Bethe-Peierls approximation, we can obtain the following expressions similar to Eq.~(\ref{eq:qimarginal}) for the cavity probabilities $q_{j\rightarrow i}^{h_j}$ and $q_{j\rightarrow k}^{h_j}$:
\begin{subequations}
  \label{eq:BP}
  \begin{align}
    q_{j\rightarrow i}^{h_j} & = \frac{1}{z_{j\rightarrow i}} e^{-\beta C(h_j)}
    w_{j\rightarrow i}(h_j) \; ,  \\
    q_{j\rightarrow k}^{h_j} & = \frac{1}{z_{j\rightarrow k}} e^{-\beta C(h_j)}
    w_{j\rightarrow k}(h_j) \; ,
  \end{align}
\end{subequations}
where the cavity statistical weights $w_{j\rightarrow i}(h_j)$ and $w_{j\rightarrow k}(h_k)$ are evaluted by
\begin{subequations}
  \label{eq:cavitywj}
  \begin{align}
    w_{j\rightarrow i}(h_j) & \equiv 
    \prod\limits_{i^\prime \in p(j)\backslash i} \Bigl[
      e^{- \beta} + (1- e^{- \beta}) \sum\limits_{h_{i^\prime}= h_j}^{D}
      q_{i^\prime \rightarrow j}^{h_{i^\prime}} \Bigr]
    \prod\limits_{k \in c(j)}
    \Bigl[e^{- \beta}+(1-e^{-\beta})\sum\limits_{h_k = 1}^{h_j}
      q_{k \rightarrow j}^{h_k} \Bigr] \; ,
    \\
    w_{j \rightarrow k}(h_j)  & \equiv
    \prod\limits_{i\in p(j)} \Bigl[e^{- \beta} + (1- e^{-\beta})
      \sum\limits_{h_i = h_j}^{D} q_{i\rightarrow j}^{h_i} \Bigr]
    \prod\limits_{k^\prime \in c(j)\backslash k} \Bigl[e^{-\beta} + (1-e^{-\beta}) 
      \sum\limits_{h_{k^\prime} = 1}^{h_j} q_{k^\prime \rightarrow j}^{h_{k^\prime}}
      \Bigr] \; ,
\end{align}
\end{subequations}
and $z_{j\rightarrow i}$ and $z_{j\rightarrow k}$ are two probability normalization constants. There are $2 M$ such self-consistent equations for a given digraph $G$ and these equations are collectively referred to as a belief propagation (BP) equation. When the inverse temperature $\beta$ is low, a fixed-point solution of the BP equation (\ref{eq:BP}) can easily be obtained by iteration. 

When $\beta$ is large we find that some of the probability values $q_j^{h_j}$ and $q_{j\rightarrow j^\prime}^{h_j}$ become very close to zero. To increase accuracy, in the numerical computations we always represent the probability values in the exponential form:
\begin{equation}
  q_j^{h_j} \equiv \exp\bigl( - u_{j}^{h_j} \bigr) \; , \quad
  {\rm and} \quad 
  q_{j\rightarrow j^\prime}^{h_j} \equiv
  \exp\bigl(- u_{j\rightarrow j^\prime}^{h_j}\bigr)
  \; . 
\end{equation}
The self-consistent equations for the positive parameters $u_{j}^{h_j}$ and $u_{j\rightarrow j^\prime}^{h_j}$ can be obtained straightforwardly from  Eq.~(\ref{eq:qimarginal}) and Eq.~(\ref{eq:BP}). At any finite $\beta$ value we find that all the coefficients $u_{j}^{h_j}$ and $u_{j\rightarrow j^\prime}^{h_j}$ keep to be bounded during the BP iteration process, but in the limiting case of $\beta =\infty$ some of these coefficients keep increasing and finally overflow. For numerical stability reasons we restrict $\beta$ to be finite in the present work.

\subsection{Thermodynamic quantities}

The probability $\rho_{i j}$ of an arc $(i, j)$ being a feedback arc is simply the probability of $h_i < h_j$. Assuming that vertices $i$ and $j$ are mutually independent in the absence of the arc $(i, j)$, we obtain
\begin{equation}
  \rho_{i j} = \frac{
    e^{-\beta}\sum\limits_{h_i= 1}^{D-1} q_{i\rightarrow j}^{h_i}
    \sum\limits_{h_j=h_i+1}^{D} q_{j\rightarrow i}^{h_j}}
      {e^{-\beta} + (1-e^{-\beta}) \sum\limits_{h_i=1}^{D} q_{i\rightarrow j}^{h_i}
        \sum\limits_{h_j=1}^{h_i} q_{j\rightarrow i}^{h_j}} \; .
\end{equation}
The average value $\rho$ of the fraction of feedback arcs in the digraph is then
\begin{equation}
  \rho = \frac{1}{M} \sum\limits_{(i, j) \in G} \rho_{i j} \; .
\end{equation}
The mean value $\varepsilon$ of the total energy density $E(\underline{h})/N$ of the system is then
\begin{equation}
  \varepsilon  =  \frac{1}{N} \Bigl[ \sum\limits_{h=1}^{D} C(h)
    \sum\limits_{i=1}^{N} q_i^{h} 
    + \sum\limits_{(j, k) \in G} \rho_{j k} \Bigr]
  = \frac{1}{D} \sum\limits_{h=1}^{D} C(h) + \alpha \rho \; .
  \label{eq:epsilon}
\end{equation}
Notice that the last equality of Eq.~(\ref{eq:epsilon}) only holds under the uniformity condition (\ref{eq:hcondition}).
 
The total free energy $F$ of the system is related to the partition function by the definition $F\equiv -\frac{1}{\beta} \ln Z$.  Under the Bethe-Peierls approximation (or through the digraph loop expansion derivation), the total free energy can be expressed as
\begin{equation}
  \label{eq:Fmodel}
  F = \sum\limits_{j=1}^{N} f_j - \sum\limits_{(j, k) \in G} f_{j k} \; ,
\end{equation}
where $f_j$ is the free energy contribution of vertex $j$ and $f_{j k}$ is the free energy contribution of arc $(j, k)$. The first term of the above expression is the total contribution of all the vertices, while the second term is the total contribution from all the arcs. The explicit expressions for $f_j$ and $f_{j k}$ are
\begin{subequations}
  \label{eq:fvandarc}
  \begin{align}
    f_j & = -\frac{1}{\beta} \ln \biggl\{ \sum\limits_{h_j=1}^{D}
    e^{- \beta C(h_j)}
    \prod\limits_{i\in p(j)}
    \Bigl[e^{-\beta} + (1- e^{- \beta}) \sum\limits_{h_i = h_j}^{D}
      q_{i\rightarrow j}^{h_i} \Bigr]
    \prod\limits_{k \in c(j)}
    \Bigl[e^{-\beta}+(1-e^{-\beta}) \sum\limits_{h_k=1}^{h_j}
      q_{k \rightarrow j}^{h_k} \Bigr] \biggr\} \; , 
    \\
    f_{j k} & = -\frac{1}{\beta} \ln \biggl\{ e^{-\beta} + (1 - e^{-\beta})
    \sum\limits_{h_j =1}^{D} q_{j\rightarrow k}^{h_j}
    \sum\limits_{h_k=1}^{h_j}
    q_{k\rightarrow j}^{h_k} \biggr\} \; .
  \end{align}
\end{subequations}
Notice the contributions of all the attached arcs of vertex $j$ are considered in computing the free energy contribution $f_j$. This is the physical reason why the free energy contributions from all the arcs should be subtracted in Eq.~(\ref{eq:Fmodel}). The free energy density $f$ is simply $f \equiv \frac{F}{N}$. The entropy density $s$ of the system is
\begin{equation}
  s  \equiv  \beta \bigl( \varepsilon - f\bigr)
  =
  \frac{1}{D}\sum\limits_{h=1}^{N} \beta C(h) + \beta ( \alpha \rho - f)
  \; .
\end{equation}

\subsection{Adjustment of the cost function $C(h)$}

We need to adjust the layer cost function $C(h)$ at each step of the BP iteration process to guarantee the uniformity condition (\ref{eq:hcondition}). Here we follow the recipe that has already been used in \cite{Sulc-Zdeborova-2010}.  Suppose at the end of the $t$-th BP iteration the computed cavity distributions are $q_{j\rightarrow k}^{h_j}(t)$ and $q_{j\rightarrow i}^{h_j}(t)$. At the $(t+1)$-th BP iteration step, these cavity distributions are taken as inputs of Eq.~(\ref{eq:wj}) to compute the weights $w_j(h_j)$ of all the $N$ vertices $j$. The probability of vertex $j$ being at layer $h$ is then $q_j^{h} \propto e^{-\beta C(h)} w_j(h)$. Because of the uniformity condition (\ref{eq:hcondition}), the layer cost function $C(h)$ needs to satisfy the following self-consistent equation
\begin{equation}
  \label{eq:Ccondition}
  e^{-\beta C(h)} = 
  \frac{N}{D\sum\limits_{j=1}^{N} \frac{w_j(h)}
    {\sum\limits_{h^{\prime}=1}^D e^{-\beta C(h^{\prime})} w_j(h^\prime)}}
  \; .
\end{equation} 
In the numerical computation, the constant term of the occupation cost function $C(h)$ is fixed by the following normalization condition 
\begin{equation}
  \label{eq:Cnorm}
  \sum\limits_{h=1}^{D} e^{-\beta C(h)} = 1 \; .
\end{equation}
Many different methods are conceivable to solve Eq.~(\ref{eq:Ccondition}) under the restriction (\ref{eq:Cnorm}). Here we adopt a very simple iterative process as follows.  A cost function $C(h)$ is fed into the right-hand side of Eq.~(\ref{eq:Ccondition}) to generate a new cost function; a constant term is then added to this output function to make it satisfy Eq.~(\ref{eq:Cnorm}); and the resulting updated cost function is then fed into the right-hand side of Eq.~(\ref{eq:Ccondition}) to start another round of refinement. We set the maximum number of iteration steps to be $50$ in the numerical code, but usually a fixed-point solution $C(h)$ is already reached within a few iteration steps.

After the cost function $C(h)$ has been adjusted by the above-mentioned method, the cavity distributions of the $(t+1)$-th BP step are then obtained through Eq.~(\ref{eq:BP}) using the cavity distributions of the $t$-th BP step and the updated function $C(h)$ as inputs.

\section{Theoretical results on three random digraph ensembles}
\label{sec:result}

We now apply the RS mean field theory to three different types of random digraph ensembles. For each digraph ensemble, we determine the minimum fraction of feedback arcs for the $D$-segmentation problem by solving the RS mean field equations through the conventional population dynamics simulation technique \cite{Mezard-Montanari-2009,Zhou-2015}. Three types of random digraphs are considered in the present paper:
\begin{enumerate}
\item[(a)]
  BRR: balanced regular random digraph. Such a digraph is also referred to as a random Eulerian digraph \cite{Gupte-etal-2011}. It is a maximally random digraph with the constraint that each vertex has $\alpha$ in-coming arcs and $\alpha$ out-going arcs (here $\alpha$ must be an integer), namely $|p(j)|=|c(j)|=\alpha$ for each vertex $j\in [1,N]$.
\item[(b)]
  RR: regular random digraph. It is a maximally random digraph with the constraint that each vertex is attached the same number ($= 2 \alpha$) of arcs, namely $|p(j)| + |c(j)| = 2 \alpha$ for each vertex $j \in [1, N]$. The direction of each arc in the digraph is chosen completely at random. 
\item[(c)]
  ER: Erd\"os-R\'enyi digraph. The $M = \alpha N$ of the digraph are added in a sequential manner. The direction of each newly added arc is chosen completely at random, and the two end points of this arc are also chosen uniformly at random from all the $N$ vertices.
\end{enumerate}

\subsection{The $D$-segmentation problem}
\label{subsec:Dseg}

\begin{figure}[t]
  \begin{center}
    \includegraphics[angle=270,width=0.8\linewidth]{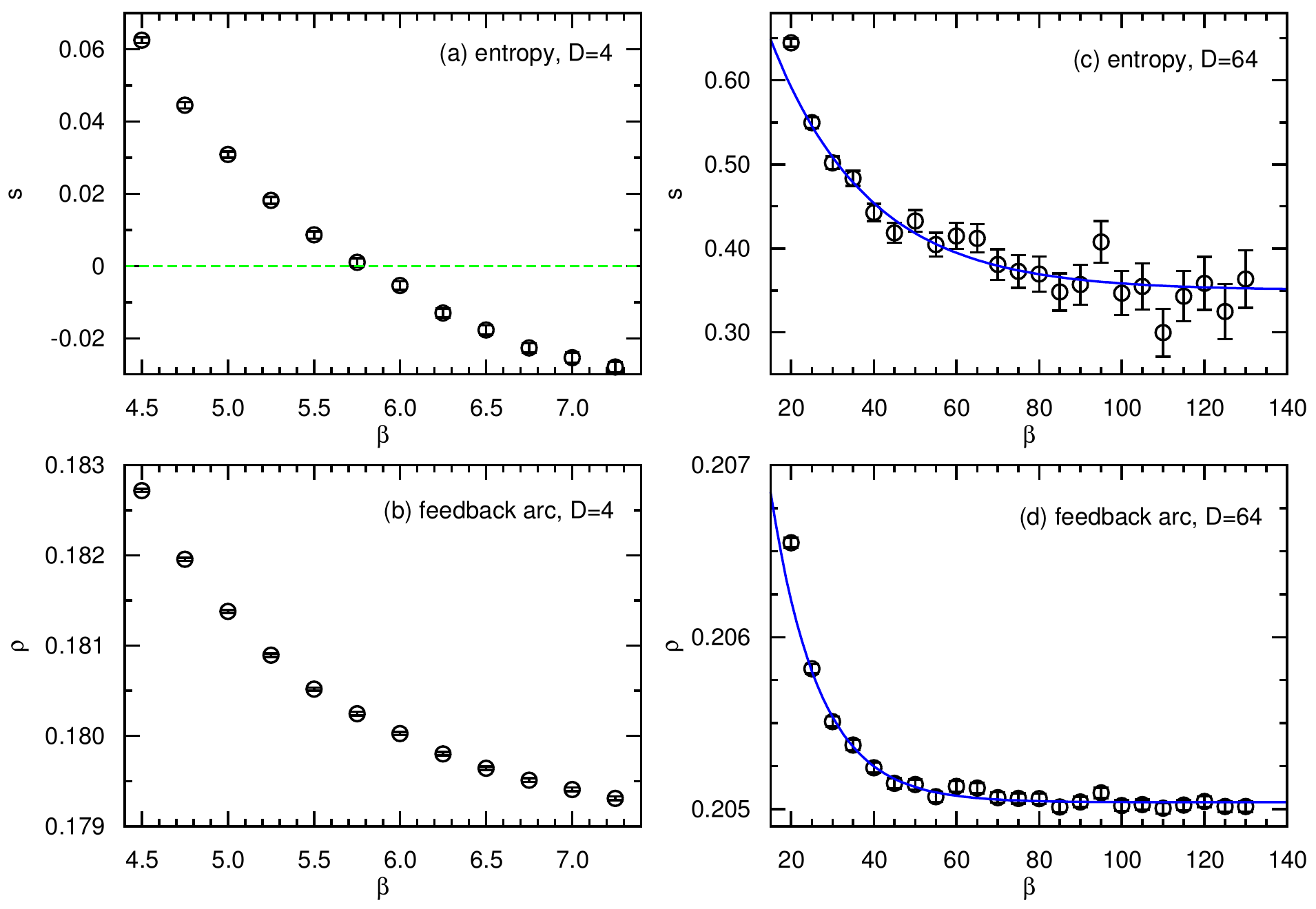}
  \end{center}
  \caption{
    \label{fig:BRRa5RS}
    Replica-symmetric theoretical results for the $D$-segmentation problem, obtained on the BRR digraph ensemble of arc density $\alpha=5$. At each value of the inverse temperature $\beta$, the predicted entropy density $s$ is shown in (a) for $D=4$ and in (c) for $D=64$, and the corresponding value of the feedback arc density $\rho$ is shown in (b) and (d). At $D=4$ the entropy density becomes negative at $\beta > 5.746$, while at $D=64$ the entropy density converges to a positive limiting value at $\beta \gg 1$. The fitting curves of (b) and (d) are $s(\beta)=0.3501 + 0.5633\ e^{-0.0423 \beta}$ and $\rho(\beta) = 0.2050 + 0.0066\ e^{-0.0861 \beta}$ (the data can also be well fitted by power-law functions $s(\beta) = 0.32 + 4.0/\beta + 52/\beta^2$ and $\rho(\beta) = 0.2051 - 0.019/\beta + 0.93/ \beta^2$, but we prefer the exponential fitting as it offers a characteristic inverse temperature).
  }
\end{figure}

As a concrete example, we show in Fig.~\ref{fig:BRRa5RS} the RS mean field results obtained on the BRR ensemble of arc density $\alpha = 5$. At each fixed layer number $D$ the entropy density $s$ and the mean feedback arc density $\rho$ both decrease with the inverse temperature $\beta$. If the digraph is divided only into $D=4$ layers (Fig.~\ref{fig:BRRa5RS}, left panel), we notice that the entropy density $s$ changes from being positive to being negative at $\beta \approx 5.746$, and the predicted feedback arc density at this point is $\rho =\rho_0 \approx 0.1803$. Since only non-negative values of entropy density are meaningful, we take  $\rho_0$ as the minimum feedback arc density for the $4$-segmentation problem (this same criterion was adopted earlier in \cite{Zhou-2013} for the undirected feedback vertex set problem and it produced results that are in agreement with the results obtained by more rigorous methods \cite{Bau-Wormald-Zhou-2002}). We therefore predict that, for a random BRR digraph of arc density $\alpha=5$, at least $18.03 \%$ of all the arcs must be pointing from lower layers to higher layers in any $4$-segmentation solution. As the ground-state entropy density is predicted to be zero, the total number of the optimal $4$-segmentation solutions will be finite (or increase at most sub-exponentially with $N$).

At a larger layer number $D=64$, the RS mean field theory still predicts that both $s$ and $\rho$ decrease with $\beta$ (Fig.~\ref{fig:BRRa5RS}, right panel), but the entropy density $s$ converges to a positive value of $s = s_0 \approx 0.3501$ as $\beta$ becomes very large. The feedback arc density also approaches a limiting value of $\rho=\rho_0 \approx 0.2050$ at large $\beta$ values. These theoretical results indicate that the ground-state entropy density for this $64$-segmentation problem is positive, and there is an exponential number ($\sim e^{s_0 N}$) of $64$-segmentation solutions with minimum feedback arc density $\rho_0$.

The minimum feedback arc density ($\rho=\rho_0$) and the ground-state entropy density ($s=s_0$) of the $D$-segmentation problem for other random digraph ensembles can be computed in the same way. For example, $\rho_0  \approx 0.1141$ at $D=4$ and $\rho_0 \approx 0.1280$ at $D=64$ for the RR digraph ensemble of arc density $\alpha=5$; and for the ER digraph ensemble of arc density $\alpha=5$, the minimum feedback arc density is $\rho_0 \approx 0.1172$ at $D=4$ and $\rho_0 \approx 0.1324$ at $D=64$.

\subsection{The minimum feedback arc set problem}

\begin{figure}[t]
  \begin{center}
    \includegraphics[angle=270,width=1.0\textwidth]{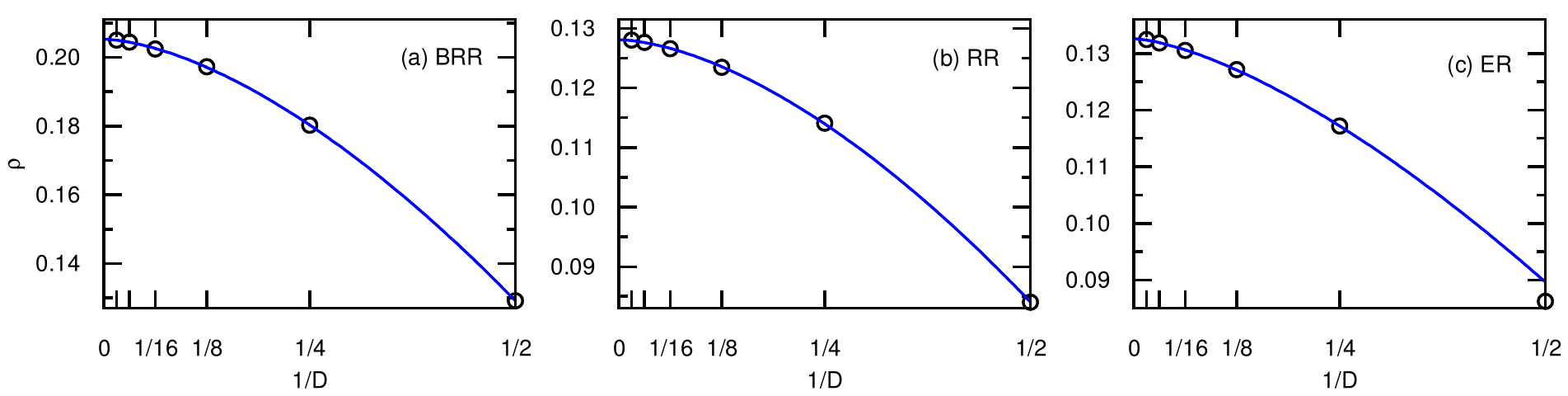}
  \end{center}
  \caption{
    \label{fig:A5RSlimit}
    Replica-symmetric theoretical results for the minimum feedback arc density of the $D$-segmentation problem, obtained on three random digraph ensembles of arc density $\alpha=5$. The theoretical data are fitted by the curve $\rho(D)= \rho_\infty - \frac{c}{D^\gamma}$ with the fitting parameters (a) $\rho_\infty = 0.2053 \pm 0.0001$, $c=0.232 \pm 0.002$, $\gamma = 1.61 \pm 0.01$ for the BRR ensemble; (b) $\rho_\infty=0.1281 \pm 0.0001$, $c=0.137 \pm 0.001$, $\gamma=1.64 \pm 0.01$ for the RR ensemble; and (c) $\rho_\infty = 0.1326 \pm 0.001$, $c=0.120 \pm 0.006$, $\gamma=1.48 \pm 0.04$ for the ER ensemble. The limiting value $\rho_\infty$ is taken as the predicted minimum feedback arc density for the minimum FAS problem.
  }
\end{figure}

As expected, the minimum feedback arc density $\rho=\rho_0$ of the $D$-segmentation problem increases with layer number $D$ for a given random digraph ensemble (see Fig.~\ref{fig:A5RSlimit}). The minimum value $\rho_0$ as a function of $D$ can be fitted nicely by the function
\begin{equation}
  \rho_0(D) = \rho_\infty - \frac{c}{D^\gamma} \; ,
  \label{eq:rhofit}
\end{equation}
where $\rho_\infty$, $c$, and $\gamma$ are three fitting parameters. Empirically we find that the decay exponent $\gamma$ takes value equal to or greater than $3/2$, therefore the minimum value of $\rho$ converges quickly with $D$. Because the $D$-segmentation problem becomes the minimum FAS problem at large $D$ values, we take the limiting value $\rho_\infty$ of Eq.~(\ref{eq:rhofit}) as the predicted minimum feedback arc density of the FAS problem. For the BRR digraph ensemble of $\alpha=5$, the numerical value is $\rho_\infty \approx 0.2053$.

Our theoretical procedure offers a convenient way of computing the ensemble-averaged minimum feedback arc density. Table~\ref{tab:fasRS} records all the numerical values obtained by this method for the three types of digraph ensembles. At each arc density $\alpha$, we notice that the value of $\rho_\infty$ for the BRR ensemble is larger than that of the ER ensemble, which is again larger than that of the RR ensemble.

Although the minimum feedback vertex set problem on undirected RR graphs has been successfully treated by rigorous probabilistic methods \cite{Bau-Wormald-Zhou-2002,Haxell-etal-2008}, according to our knowledge, no tight mathematical bound on the minimum feedback arc density of random digraph ensembles is available in the literature. We are therefore unable to compare the theoretical results of Table~\ref{tab:fasRS} with rigorous mathematical results. However, we find that the RS results are consistent with the results obtained by the simulated annealing (SA) algorithm of \cite{Zhao-Zhou-2016b}. For example, at arc density $\alpha=5$, the SA algorithm achieves many FAS solutions of arc density $\rho\approx 0.2212$ for a single BRR digraph of $N=10^5$ vertices, which is slightly exceeding the predicted minimum value of $0.2053$; the SA algorithm achieves many solutions of $\rho \approx 0.1409$ for an ER digraph of $N=10^5$, which is again slightly beyond the predicted value of $0.1326$.

\begin{table}[t]
  \caption{\label{tab:fasRS}
    Replica-symmetric theoretical results on the minimum feedback arc set problem. The first column is the arc density $\alpha$ of the random digraph ensemble. The RS mean field predictions for the minimum fraction of feedback arcs are recorded in the second column (for BRR digraphs), the third column (for RR digraphs), and the fourth column (for ER digraphs).}
  \begin{center}
    \begin{tabular}{r|c|c|c}
      \hline \hline 
      $\alpha$ \  & \ \ \ BRR \ \ \  & \ \ \ RR \ \ \      & \ \ \ ER \ \ \ \\
      \hline
      $1$  \   & $0$      & $0$      & $0$          \\
      $2$  \   & $0.0939$ & $0.0130$ & $0.0266$     \\
      $3$  \   & $0.1456$ & $0.0590$ & $0.0687$     \\
      $4$  \   & $0.1800$ & $0.0979$ & $0.1039$     \\
      $5$  \   & $0.2053$ & $0.1281$ & $0.1326$     \\
      $6$  \   & $0.2250$ & $0.1524$ & $0.1563$     \\
      $7$  \   & $0.2411$ & $0.1724$ & $0.1755$     \\
      $8$  \   & $0.2547$ & $0.1893$ & $0.1918$     \\
      $9$  \   & $0.2661$ & $0.2037$ & $0.2058$     \\
      $10$ \   & $0.2760$ & $0.2163$ & $0.2180$     \\
      \hline \hline
    \end{tabular}
  \end{center}
\end{table}

\section{Algorithmic applications}
\label{sec:algorithm}

The RS mean field theory, besides being a powerful tool for revealing ensemble-averaged properties of random digraphs, is also a major source of inspiration for heuristic algorithms. Here we describe two simplest ways of integrating the BP equation into an efficient solver for the $D$-segmentation problem and the minimum FAS problem.

\subsection{Belief propagation guided decimation (BPD)}

The basic idea of the BPD algorithm is straightforward: the vertices $i$ of an input digraph $G$ are fixed to different layers $h$ according to their marginal distributions $q_i^{h}$ to simplify the optimization problem  \cite{Mezard-etal-2002}. In our implementation, we first iterate the BP equation (\ref{eq:BP}) in combination with the condition (\ref{eq:Ccondition}) a few number $t_0$ of times  to drive the probability distributions close to a fixed point ($t_0 = 100$ in our code). Then we fill the $D$ different layers up to their capacity as determined by Eq.~(\ref{eq:uniformity}), starting from the layer at the top ($h=D$) and proceeding downward to the layer at the bottom ($h=1$). (We adopt this particular order of filling mainly for the convenience of implementing the code. Other recipes might be even better in performance but we have not yet tested any of them.) Suppose at the beginning of the $r$-th ($r=1, 2, \ldots$) decimation step the layer $h=H$ can still accommodate $n(H)$ more vertices while all the layers $h^\prime > H$ are no longer available. We rank in descending order the remaining un-assigned vertices $i$ according to their marginal probability value $q_{i}^{H}$, and then assign the first $n_0$ (e.g., $n_0=10^{-3} N$)  vertices to layer $H$  (in the special situation of $n_0 > n(H)$, only the first $n(H)$ vertices are assigned to layer $H$). To finish the $r$-th decimation step, we then iterate the BP equation (\ref{eq:BP}) in combination with Eq.~(\ref{eq:Ccondition}) a few number $t_1$ of times ($t_1 = 10$) and re-evaluate the marginal probability distributions of the remaining un-assigned vertices according to Eq.~(\ref{eq:qimarginal}) under the addition restriction of layer state $h\leq H$.

\begin{table}[t]
\caption{\label{tab:BPD}
  Solving the $D$-segmentation problem (with $D=100$) by the BPD algorithm. Three digraph instances of size $N=10^5$ and arc density $\alpha=5$ are considered: $G_{BRR}$ is a balanced regular random digraph, $G_{RR}$ is a regular random digraph, and $G_{ER}$ is an Erd\"os-R\'enyi digraph. At each inverse temperature ranging from $\beta=6.0$ to $\beta=24.0$ the BPD algorithm is repeated $40$ times with independent random number seed and the density of feedback arcs in the best $100$-segmentation solution is recorded here (second to eleventh columns).}
\vskip 0.1cm
\begin{center}
  \begin{tabular}{l|cccccccccc}
    \hline \hline 
    $\beta$   & $6.0$     & $8.0$     & $10.0$   & $12.0$   & $14.0$   & $16.0$    & $18.0$   & $20.0$    & $22.0$    & $24.0$ \\
    \hline
    $G_{BRR}$  & $0.2302$  & $0.2293$  & $0.2292$ & $0.2292$ & $0.2292$ & $0.2294$ & $0.2296$ & $0.2298$  &           &           \\
    $G_{RR}$   & $0.1435$  & $0.1404$  & $0.1388$ & $0.1379$ & $0.1374$ & $0.1371$ & $0.1370$ & $0.1369$  & $0.1369$  & $0.1368$ \\
    $G_{ER}$   & $0.1485$  & $0.1454$  & $0.1440$ & $0.1432$ & $0.1428$ & $0.1425$ & $0.1422$ & $0.1425$  & $0.1423$  & $0.1425$ \\
    \hline \hline
  \end{tabular}
\end{center}
\end{table}

For the $D$-segmentation problem with $D=100$, the performance of this BPD algorithm is examined on three random digraph instances of size $N=10^5$ and arc density $\alpha=5$, see Table~\ref{tab:BPD}. We notice that the inverse temperature $\beta$ is not a sensitive parameter of the algorithm. Almost equally good solutions are obtained by BPD at different values of $\beta \in [6.0, 24.0]$. For the BRR digraph instance, the best $100$-segmentation solutions obtained by BPD have a fraction $\rho_{BPD}\approx 0.2292$ of feedback arcs, which is $11.6\%$ beyond the predicted fraction of $\rho_{RS} \approx 0.2053$ by the RS theory; for the RR instance, BPD reaches $\rho_{BPD}\approx 0.1368$, which is $6.8\%$ beyond the RS prediction of $\rho_{RS}\approx 0.1281$; for the ER instance, BPD reaches $\rho_{BPD}\approx 0.1422$, which is $7.2\%$ beyond the RS prediction of $\rho_{RS}\approx 0.1326$. 

Small but noticeable gaps between the BPD results and the RS theoretical results are also observed in other random digraph instances. This gap is more pronounced as the arc density $\alpha$ of the digraph becomes larger (i.e., the input digraph becomes more denser). A major shortcoming of the BPD algorithm is that the state $h_i$ of a vertex $i$ can not be changed once it has been assigned a value. Errors may therefore accumulate and be amplified during the BPD process. As we will see in the next subsection, these errors can be significantly reduced by allowing the vertices to tune their states through a learning process. (The backtracking procedure discussed in \cite{Marino-etal-2016} may also be capable of significantly reducing the assignment errors; we leave this as an interesting issue for future explorations.)

\subsection{Belief propagation guided reinforcement (BPR)}

The BPR algorithm is based on the idea of reinforcement learning and its first application was on the binary perceptron problem \cite{Braunstein-Zecchina-2006}. In our implementation for the $D$-segmentation problem, a non-negative memory function $\phi_j^{h}$ is introduced for each vertex $j$ of the digraph. This memory function can be interpreted as a prior probability distribution over the different layers $h \in [1, D]$, and it is iteratively refined during the learning process. After taking into account the prior distribution $\phi_j^{h_j}$ of each vertex $j$,  the expression (\ref{eq:qimarginal}) for the marginal distribution $q_j^{h_j}$ is modified as
\begin{equation}
  \label{eq:qjBPR}
  q_j^{h_j} = \frac{1}{z_j} e^{-\beta C(h_j)} \phi_j^{h_j} w_j(h_j) \; ,
\end{equation}
where the weight $w_j(h_j)$ is computed through Eq.~(\ref{eq:wj}). Accordingly, the self-consistent equation for the occupation cost $C(h)$ is changed to
\begin{equation}
  \label{eq:CconditionBPR}
  e^{-\beta C(h)} = 
  \frac{N}{D\sum\limits_{j=1}^{N} \frac{\phi_j^{h} w_j(h)}
    {\sum\limits_{h^{\prime}=1}^D e^{-\beta C(h^{\prime})} \phi_j^{h^\prime} 
      w_j(h^\prime)}}
  \; .
\end{equation}
Similar to Eq.~(\ref{eq:qjBPR}),  the BP equation (\ref{eq:BP}) for the cavity probability distributions is modified as
\begin{subequations}
  \label{eq:BPr}
  \begin{align}
    q_{j\rightarrow i}^{h_j} & = \frac{1}{z_{j\rightarrow i}} e^{-\beta C(h_j)}
    \phi_j^{h_j} w_{j\rightarrow i}(h_j) \; ,
    \\
    q_{j\rightarrow k}^{h_j} & = \frac{1}{z_{j\rightarrow k}} e^{-\beta C(h_j)}
    \phi_j^{h_j} w_{j\rightarrow k}(h_j) \; ,
  \end{align}
\end{subequations}    
where the cavity weights $w_{j\rightarrow i}(h_j)$ and $w_{j\rightarrow k}(h_j)$ are computed through Eq.~(\ref{eq:cavitywj}).

\begin{figure}
  \begin{center}
    \includegraphics[angle=270,width=1.0\textwidth]{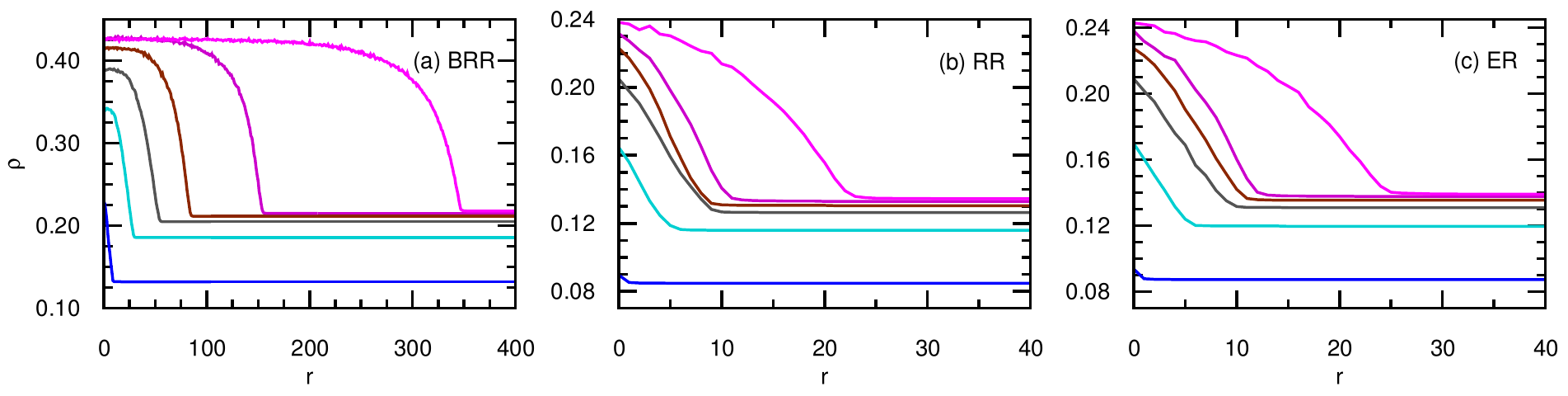}
  \end{center}
  \caption{
    \label{fig:BPRevolve}
    Evolution of the BPR learning process. The horizontal axis is the step index $r$ of the BPR process. The vertical axis is the fraction $\rho$ of feedback arcs in the $D$-segmentation solution reached at the $r$-th BPR step. Results obtained on three random digraph instances of size $N=10^5$ and arc density $\alpha=5$ are shown here: (a) the BRR instance; (b) the RR instance; and (c) the ER instance. These three digraph instances are the same as those used in Table~\ref{tab:BPD} and Table~\ref{tab:BPR}. The total layer number $D$ is set to be $D=2, 4, 8, 16, 32, 64$ (from bottom to top) for the six curves of each panel. The inverse temperature is fixed to be $\beta=10$ in all these simulations.  
  }
\end{figure}

Starting from an initial uniform prior distribution $\phi_j^{h_j}=\frac{1}{D}$ for each vertex $j$, at each evolution step $r=1,2, \ldots$ of the BPR process the following computations are carried out in a sequential order: (1) the BP equation (\ref{eq:BPr}) in combination with Eq.~(\ref{eq:CconditionBPR}) is iterated a number $t_1$ of times (we chose $t_1=10$ as in BPD); (2) then the marginal distribution $q_j^{h_j}$ of each vertex $j$ is determined according to Eq.~(\ref{eq:qjBPR}), and the value $h_j=h_j^*$ at which $q_j^{h_j}$ achieves maximum is recorded, and the memory function $\phi_j^{h_j}$ of vertex $j$ is then simply updated as $\phi_j^{h_j} \leftarrow \phi_j^{h_j}$ for $h_j \neq h_j^*$ and $\phi_j^{h_j} \leftarrow (1+\eta) \phi_j^{h_j}$ for $h_j=h_j^*$, where $\eta$ is a small learning parameter (we set $\eta=0.05$); (4) the layer configuration $\underline{h}^{(r)} \equiv (h_1^*, h_2^*, \ldots, h_N^*)$ is regarded as the candidate solution at the $r$-th BPR step. Because of the stochasticity of the learning process, this solution $\underline{h}^{(r)}$ satisfies the uniformity condition (\ref{eq:uniformity}) only approximately but not exactly. (For the minimum FAS problem this might actually be an advantage, since relaxing slightly Eq.~(\ref{eq:uniformity}) will enhance the exploration of additional low-energy configurations.)

Figure~\ref{fig:BPRevolve} demonstrates how the $D$-segmentation solutions improve with time for three random digraphs of the same size $N=10^5$ and the same arc density $\alpha=5$. For the RR and the ER digraph instances, we observe that the feedback arc fraction $\rho$ gradually decreases with the evolution step $r$ and then saturates at a final low value, see Fig.~\ref{fig:BPRevolve}(b) and \ref{fig:BPRevolve}(c). Such steadily improving patterns indicate that near-optimal solutions for the RR and the ER instances can be reached by the BPR process in a smooth way and very quickly (e.g., within $30$ evolution steps).  However, when the BPR algorithm is tested on the BRR instance, we observe that the feedback arc fraction $\rho$ may fluctuate around the initial high value for many evolution steps before the learning process reaches a tipping point (the top curve of Fig.~\ref{fig:BPRevolve}(a) for $D=64$ shows this initial struggling behavior most clearly). The existence of an initial plateau phase suggests that the BRR instance, with all the vertices having the same number of in-coming and out-going arcs, is much harder to solve and the learning process needs to accumulate more probabilistic information to figure out the correct direction of evolution.

\begin{table}[t]
  \caption{\label{tab:BPR}
    Solving the $D$-segmentation problem (with $D=100$) by BPR and the minimum feedback arc set problem by nested-BPR. The same three digraph instances $G_{BRR}$, $G_{RR}$ and $G_{ER}$ of Table~\ref{tab:BPD} are considered here (size $N=10^5$, arc density $\alpha=5$). The second column records the feedback arc density in the $100$-segmentation solution obtained by a single run of the BPR algorithm, the third to sixth columns record the feedback arc density in the FAS solution obtained by a single run of the nested-BPD algorithm with fixed division number $D=4$, $8$, $16$, or $100$. The seventh column records the feedback arc density in the FAS solution obtained by a single run of the simulated annealing algorithm \cite{Zhao-Zhou-2016b}. The inverse temperature is fixed to $\beta=10$ in the BPR and nested-BPR algorithms.
  }
  \vskip 0.1cm
  \begin{center}
    \begin{tabular}{l|ccccccccccc}
      \hline \hline 
      & $100$-segmentation  & FAS ($D=4$)    & FAS ($D=8$) & FAS ($D=16$) & FAS ($D=100$) & FAS (SA) \\
      \hline
      $G_{BRR}$  & $0.2194$            & $0.2362$       & $0.2245$    & $0.2202$    & $0.2196$  & $0.2212$ \\
      $G_{RR}$   & $0.1348$            & $0.1443$       & $0.1383$    & $0.1357$    & $0.1349$  & $0.1352$ \\
      $G_{ER}$   & $0.1404$            & $0.1501$       & $0.1438$    & $0.1411$    & $0.1404$  & $0.1409$ \\
      \hline \hline
    \end{tabular}
  \end{center}
\end{table}

For the $D$-segmentation problem with $D=100$, the final fractions $\rho$ of feedback arcs reached by a single run of the BPR algorithm on the same three digraph instances used in the preceding subsection are listed in the second column of Table~\ref{tab:BPR}. These results are noticeably lower than the corresponding results of Table~\ref{tab:BPD} obtained by the BPD algorithm. Besides the ability of reaching better solutions, the BPR algorithm is also faster than the BPD algorithm.

\subsection{The nested-BPR (nBPR) algorithm}

After a $D$-segmentation solution is constructed by the above-mentioned BPR (or BPD) algorithm, we can delete all the upward (feedback) arcs to break all the directed cycles connecting vertices of different layers. However, directed cycles may still exist within each single layer of the solution, and the number of such ``internal'' cycles may still be huge. To completely break all the directed cycles in the input digraph, a natural idea is to apply the BPR process on the sub-digraph induced by all the vertices of the same layer and divide it again into $D$ sub-layers. This BPR process can be recursively applied on each newly formed sub-layer to further divide it into smaller parts (see Fig.~\ref{fig:multiple}). After this nested BPR (nBPR) process is finished and all the upward arcs are deleted, the remaining digraph will be free of any directed cycles. The set formed by all the upward arcs then must be a feedback arc set for the input digraph.

The results obtained by this nBPR algorithm on three digraph instances are listed in Table~\ref{tab:BPR} (third to sixth columns). Compared with the results obtained by simulated annealing (seventh column of Table~\ref{tab:BPR}), we are very satisfied to observe that nBPR with $D=16$ already performs better. Another big advantage of the nBPR algorithm is its speed. For example at $D=16$, nBPR running on a computer of frequency $2.5$ Ghz reaches in less than $4.4$ hours a close-to-minimum FAS solution of arc density $\rho = 0.2202$ for the difficult BRR digraph instance ($N=10^5$ and $\alpha=5$), while it takes the SA algorithm more than $63.1$ hours to reach a final FAS solution of larger arc density $\rho = 0.2213$. The reason behind the high speed of nBPR is that each vertex has only a small number $D \sim 10$ of different states during the whole search process. There is a high degree of flexibility in implementing the nBPR algorithm, it can easily be made much faster to tackle truly big digraphs with billions of vertices and arcs (e.g., by setting $D=4$ or even $D=2$).

\section{Conclusion and outlook}
\label{sec:conclusion}

As a brief summary, we tackled the minimum feedback arc set problem by first generalizing it to the digraph $D$-segmentation problem and then solving the latter by the replica-symmetric mean field theory and its associated message-passing algorithms. Our theoretical approach enables us to predict the minimum FAS cardinality for different random digraph ensembles, and the divide-and-conquer nested-BPR (or nested-BPD) algorithm greatly reduces the search time needed to reach close-to-minimum FAS solutions. This work, as a significant step beyond our previous efforts in \cite{Zhao-Zhou-2016b}, will likely be useful in future theoretical and algorithmic researches on directed graphs.

As demonstrated in Fig.~\ref{fig:BRRa5RS}(a), the entropy density predicated by the RS mean field theory may become negative at large values of the inverse temperature $\beta$. This non-physical result is a strong indication that replica symmetry is spontaneously broken. Furthermore, the fact that the BPR algorithm gives good solutions for the $D$-segmentation problem might suggest that the energy landscape of this problem is highly nontrivial. Very likely the ground states and low-energy states of the $D$-segmentation problem (and the FAS problem) form many distinct clusters which are separated by high energy barriers. To better understand the solution space structure of these cycle-constrained hard problems and to improve the theoretical prediction on the minimum number of feedback arcs, we need to carry out the more challenging first-step replica-symmetry-breaking (1RSB) mean field calculations \cite{Mezard-Parisi-2001,Mezard-Montanari-2009}. To gain some initial experiences, especially on how to treat the global constraint (\ref{eq:uniformity}) within the 1RSB theory, one may start with the zero-temperature limit ($\beta = \infty$) and a small value of layer number $D$.

Rigorous probabilistic analysis on the $D$-segmentation problem is yet to be accomplished. This task is likely to be very hard for general random digraphs, but it may be feasible to derive tight lower bounds of feedback arc density for the simplest BRR digraph ensemble. The methods developed in \cite{Bau-Wormald-Zhou-2002} for the undirected feedback vertex set problem might offer some inspiration for handling the $D$-segmentation problem. We will be very happy to know any new progress along this direction.

It may also be very helpful to map the directed feedback vertex set problem to a similar $D$-segmentation problem. We will explore this idea and carry out theoretical and algorithmic investigations to improve the results documented in \cite{Zhou-2016b}.

The concepts of feedback arcs and feedback vertices are closely related to the structural and dynamical properties of directed networks. Efficient and accurate heuristic algorithms are highly desirable in real-world applications, especially in studying huge social networks with billions of vertices and arcs. We expect the nested-BPR and nested-BPD algorithms to be highly competent for large-scale network analysis. These message-passing algorithms can also be used in combination with other heuristic algorithms such as simulated annealing \cite{Zhao-Zhou-2016b,Galinier-Lemamou-Bouzidi-2013}.

\section*{Acknowledgement}

We thank Dr. Jin-Hua Zhao for an earlier collaboration which stimulated the present project, and thank Dr. Heiko Bauke for a helpful correspondence on the TRNG library of random number generators \cite{Bauke-Mertens-2007} which was called in our computer simulations. One of the authors (HJZ) acknowledges the hospitality of the Asia Pacific Center for Theoretical Physics (APCTP, Pohang, Korea) where the theoretical part of this project was carried out during a short visit in November 2016. This research was partially supported by the National Natural Science Foundation of China (grant numbers 11121403 and 11647601).



\end{document}